\documentclass[aps,prc,twocolumn,superscriptaddress,nofootinbib]{revtex4-2}
\usepackage{graphicx,amsfonts,color,epsfig,epstopdf}
\usepackage{footnote,multirow}
\usepackage{amsmath}
\usepackage{amssymb}
\usepackage{physics}
\usepackage[utf8]{inputenc}

\newcommand{\eg}{\emph{e.g.}}

\begin{document}
\title{Microscopic optical potentials from a Green's function approach}
\author{G. H. Sargsyan}
\affiliation{Lawrence Livermore National  Laboratory, Livermore, California 94550, USA}
\affiliation{Facility for Rare Isotope Beams, Michigan State University, East Lansing, Michigan 48824, USA}
\author{G. Potel}
\affiliation{Lawrence Livermore National  Laboratory, Livermore, California 94550, USA}
\author{K. Kravvaris}
\affiliation{Lawrence Livermore National  Laboratory, Livermore, California 94550, USA}
\author{J. E. Escher}
\affiliation{Lawrence Livermore National  Laboratory, Livermore, California 94550, USA}

\begin{abstract}
Optical potentials are a standard tool in the study of nuclear reactions, as they describe the interaction between a target nucleus and a projectile. The use of phenomenological optical potentials built using experimental data on stable isotopes is widespread. Although successful in their dedicated domain, it is unclear whether these phenomenological potentials can provide reliable predictions for unstable isotopes. To address this problem, optical potentials based on microscopic nuclear structure input calculations prove to be crucial, and are an important current line of research. In this work, we present an explicit implementation of the Feshbach formalism for the systematic derivation of optical potentials using input from nuclear structure models. Numerical tools for the derivation of Green's functions associated with non-local potentials are presented. The new optical potential, based on the valence shell model, is applied to the calculations of $n+^{24}$Mg elastic scattering and yields a close agreement with the experimental data.

\end{abstract}

\maketitle
\section{Introduction}
Nuclear reactions serve a twofold purpose in nuclear physics studies. On the one hand, they are an essential experimental tool to probe our knowledge of nuclear structure in accelerated beam facilities. On the other hand, they are at the basis of applications in nuclear astrophysics, medicine, energy, industry, and national security \cite{LRP2023,brown2024motivations}. The experimental availability of unstable nuclear species in radioactive beam facilities, and the relevance of reaction rates involving nuclei away from stability, underscores the need for a quantitatively accurate description of reaction processes in exotic isotopes. In particular, it is pressing to integrate recent advances in nuclear structure, applied to nuclei far from the valley of stability, and the quest for a unified description of the structure and reaction aspects of nuclear collisions is a highly active area of research. 

In this context, the community widely recognizes the essential role of {the optical potential (OP), which describes the interaction of a nucleon with a target nucleus accounting for its composite nature} \cite{HebbornNPDH2022,ThompsonN2009, Dickhoff:19}. OPs are essential ingredients in the analyses of direct reactions, such as elastic and inelastic scattering, transfer reactions, breakup and knockout reactions, etc. They are also necessary ingredients  to describe the interaction between the reaction constituents both in the entrance and exit channels in the compound nucleus reactions, allowing for the calculation of transmission coefficients in the Hauser-Feshbach statistical model \cite{ThompsonN2009}.  Radioactive-beam facilities explore nuclei far from stability, and modeling rare-isotope reactions requires accurate OPs, which may challenge existing models.

Non-dispersive \cite{BecchettiG1969,Perey:76,VarnerTMLC1991,KoningD2003,PruittER2023} and dispersive \cite{MorillonR2004,MorillonR2007} phenomenological OPs fitted on stable isotopes to elastic scattering and other reaction data are very successful in their dedicated range of applicability. Observables such as charge densities and single particle spectra can be added to the body of fitted experimental data in order to derive phenomenological dispersive OPs making  use of Green's function approaches \cite{MahauxS1991,MahzoonCDDW2014,AtkinsonD2019,Dickhoff:19}. Although phenomenological dispersive OPs allow for a connection with the nuclear many-body structure through the Dyson equation (see, e.g., \cite{Dickhoff:05,Dickhoff:19}), it is unclear whether they will provide reliable predictions for unstable isotopes such as those that play an important role in reaction networks relevant for astrophysical and societal applications. Some of these isotopes are experimentally available in radioactive beam facilities. Therefore, a  more explicit and direct link with the underlying structure is desirable, and even essential in regions of the nuclear chart where experimental information is scarce or non-existent \cite{HebbornNPDH2022}.

There is a very large  body of literature addressing the connection of the OP with the underlying nuclear structure. A class of microscopic OPs makes use of a  local density approximation in order to fold a one-body density with a  nucleon-nucleon interaction based on nuclear matter calculations \cite{JeukenneLM1977, BaugeDG1998}. Recent developments within this approach make use of modern chiral interactions \cite{WhiteheadLH2019}. For energies above $\sim$40 MeV g-matrix methods are commonly used \cite{Amos2000, ArellanoB2011}, while for energies $\gtrsim 65$ MeV, multiple scattering approaches, based on microscopic one-body densities, provide a good reproduction of elastic cross sections and polarization observables \cite{burrowsInitioLeadingOrder2020,BakerEDL2024,VorabbiBSFG2024}. 
Furthermore, {several structure-based methods have excelled at describing nuclear scattering at lower energies ($<$50 MeV).} The so-called Nuclear Structure Model has been successful at describing scattering observables using random-phase approximation {(RPA)} calculations on closed-shell nuclei \cite{vinh1970theory,BlanchonDAV2015}. {Quasiparticle} RPA was used in conjunction with a folding model with large-scale coupled-channel calculations to describe inelastic scattering cross-sections \cite{Nobre:10,Nobre:11}. \textit{Ab initio} calculations of OPs have been implemented in the context of the self-consistent Green's function \cite{IdiniBN2019}, Coupled-Cluster \cite{RotureauDHNP2017} and Symmetry-Adapted No-Core Shell Model \cite{BurrowsLMBSDL2024} approaches.
In such models, however, achieving the required absorption has proven to be challenging \cite{RotureauDHNP2017}.  A more detailed and exhaustive comparative account of these different approaches can be found in \cite{HebbornNPDH2022}.

Feshbach provided an early formal derivation of the OP in terms of the structure of the composite system \cite{Feshbach1958}. This approach connects explicitly the OP with the underlying many-body Hamiltonian, the resulting OP being, by construction, complex and non-local {(here non-local refers to the coordinate-space representation)}, and the real and imaginary parts are connected by the Kramers-Kronig dispersion relations.
From a microscopic perspective, the nonlocality of the optical potential arises because the underlying nucleon-nucleon effective interaction is inherently non-local \cite{ArellanoB2018}. Additionally, even if we {consider a} {local but finite-range} interaction between nucleons, non-local  contributions to the OP would still emerge from the Fock exchange term and the dynamical virtual excitations of the composite many-body system which contribute to the dispersive energy dependence of the OP \cite{MahauxBBD1985}. {A more detailed discussion on causes of nonlocalities in OPs can be found in Ref. \cite{FraserAKC2008} }

In this paper, we present a framework inspired by the Feshbach approach to obtain the first (to our knowledge) OP derived from the valence shell model. {This framework has similarities with the other structure-based methods mentioned above, but uses the rich structure of the composite nucleus calculated using multiple valance shells.} We illustrate the general approach with the neutron ($n$) + $^{24}$Mg elastic scattering reaction, for incident energies up to around 8 MeV.  To benchmark our implementation {of the Feshbach formalism}, we first reproduce earlier results of proton ($p$) elastic scattering on $^{40}$Ca, obtained  within a more schematic structure model \cite{RaoRS1973}. A distinctive trait of the present approach {compared to the earlier works} is the implementation of an iterative method for the calculation of the Green's function, enforcing its consistency with the calculated OP.

\section{Theoretical method}
\subsection{Derivation of the Optical Potential}\label{SectIIA}
We follow rather closely the formulation  given by Feshbach in Ref. \cite{Feshbach1958}. The Schr\"odinger equation for  the system consisting of an incident nucleon impinging on a target  of mass A is given by,
\begin{align}\label{eq:SchA+1}
  \left( E-H_A(\xi)-V(\mathbf r_n,\xi)-T\right)\Psi(\mathbf r_n,\xi)=0.
\end{align}
 Here ${\mathbf r_n}$ is the projectile coordinate, $T$ is the incident nucleon kinetic energy {in the center of mass frame}, $H_A(\xi)$ is the target nucleus many-body Hamiltonian that satisfies $H_A(\xi)\phi_i(\xi)=\epsilon_i\phi_i(\xi)$, with $\epsilon_i$ and $\phi_{i}(\xi)$ being the eigenenergies and eigenstates of the target nucleus, respectively, $\xi$ denotes all relevant  coordinates of the $A$ nucleons in the target {and $V(\mathbf r_n,\xi)$ is the {real, energy-independent} potential of the projectile in the field of the target nucleons.} {One can} expand the wavefunction $\Psi$ in terms of a complete set of eigenstates  $\phi_{i}(\xi)$,  
 \begin{align}\label{eq:wfn}
  \Psi(\mathbf r_n,\xi)=\phi_0(\xi) \chi_0(\mathbf r_n)+\sum_{i\neq0}\phi_i(\xi)\chi_i(\mathbf r_n),
\end{align}
where $\chi_i$ are the (unknown) wavefunctions describing  the projectile states. The states with $i\neq0$ correspond to inelastic scattering. Let us now  project (\ref{eq:SchA+1}) on the nuclear intrinsic  states $\phi_i(\xi)$,  obtaining the set of coupled  equations,
\begin{align}\label{eq:coupled}
  \left(E_i-U_{ii}(\mathbf r_n)-T\right)\chi_i(\mathbf r_n)=\sum_{j\neq i}U_{ij}(\mathbf r_n)\chi_j(\mathbf r_n),\nonumber \\ \mathrm{with} \quad i,j=0,1\dots
\end{align}
where channel energies verify  $E_i= E-\epsilon_i$.
The coupling potentials  are defined as 
\begin{align}\label{eq:coupl_pot}
  U_{ij}(\mathbf r_n)=\int\phi^*_j(\xi)V(\mathbf r_n,\xi)\phi_i(\xi)\,d\xi.
\end{align}
 
 In order to obtain the optical potential, we single out the ground state of the target by rewriting the system of equations (\ref{eq:coupled}) as,
\begin{align} \label{eq:coupled_U0}
  &\left(E-U_{00}(\mathbf r_n)-T\right)\chi_0(\mathbf r_n)=\sum_{i\neq0}U_{0i}(\mathbf r_n)\chi_i(\mathbf r_n),\\
\label{eq:coupled_Uij}  &\sum_{j\neq0}\left(E_i\delta_{ij}-U_{ij}(\mathbf r_n)-T\delta_{ij}\right)\chi_j(\mathbf r_n)=U_{i0}(\mathbf r_n)\chi_0(\mathbf r_n), \\
\nonumber&\mathrm{with} \quad i \neq 0,
\end{align}
where we have set $\epsilon_{0}=0$ and, to arrive at Eq. (\ref{eq:coupled_Uij}), we have taken all the terms besides $j=0$ in the sum in Eq. (\ref{eq:coupled}) to the left hand side.
We now define the Green's function $G^{Q}$ restricted to the excited states of the target ($Q$-space), 
\begin{align}\label{eq:Gmatrix}
  G^{Q}_{ij}(E)=\lim_{{\alpha}\to0^+}\big((E_i-T+i{\alpha})\delta_{ij}-U_{ij}(\mathbf r_n)\big)^{-1},\quad i,j\neq0. 
\end{align}
Since the vanishing parameter ${\alpha}$ is positive and the coupling potentials $U_{ij}(\mathbf r)$ are real, the analytic continuation of the Green's function on the complex $E$ plane does not have poles on the upper half-plane of $E$. Eq. (\ref{eq:coupled_Uij}) can now be formally solved, 
\begin{align}\label{eq:proj_wfn}
  \chi_i(\mathbf r_n)=\sum_{j\neq0}G^{Q}_{ij}(E)\,U_{j0}(\mathbf r_n)\,\chi_0(\mathbf r_n).
\end{align}
Substituting Eq. (\ref{eq:proj_wfn}) in (\ref{eq:coupled_U0}), we obtain a decoupled  equation for the elastic channel wavefunction $\chi_{0}$,
\begin{align}\label{eq:shred}
 \nonumber &\left(T + U_{00}(\mathbf r_n) +\sum_{i,j\neq0}U_{0i}(\mathbf r_n)\,G^{Q}_{ij}(E)\,U_{j0}(\mathbf r'_n)-E\right)\\
 &\times\chi_0(\mathbf r_n)=0.
\end{align}
The non-local, complex, and energy-dependent OP is identified with the effective potential in the above equation, 
\begin{eqnarray}
\mathcal V(\mathbf{r,r'},E)&=& U_{00}(\mathbf r)+\sum_{i,j \neq 0} U_{0i}(\mathbf{r})G^{Q}_{ij}(\mathbf{r,r'},E)U_{j0}
(\mathbf{r'}) \nonumber \\
&=&U_{00}(\mathbf r)+V_\text{dpp}(\mathbf{r,r'},E),  
\label{eq:OP}
\end{eqnarray}
where we have dropped the subindex ${n}$, and have defined the dynamic polarization potential,
\begin{align}\label{eq:100}
V_\mathrm{dpp}(\mathbf{r,r'},E)=\sum_{i,j \neq 0} U_{0i}(\mathbf{r})G^{Q}_{ij}(\mathbf{r,r'},E)U_{j0}(\mathbf{r'}).
\end{align}
This dynamic polarization potential has, in general, an angular-momentum dependence determined by the multipolar decomposition of the coupling potentials and the Green's function.

Since the real coupling potentials $U_{ij}(\mathbf r)$ are finite and energy-independent the above optical potential will have the same poles as the Green's function (\ref{eq:Gmatrix}) as a function of the complex energy, and will thus be analytical in the upper half-complex plane. As a consequence, the optical potential is dispersive and the energy-dependence of the real and imaginary parts are connected through the Kramers-Kronig dispersion relations. (see \cite{Lipperheide:66}, see also Eq. (2.26) of \cite{Feshbach1958}).

The calculations presented in this work are based on an explicit implementation of Eq. (\ref{eq:OP}) in a truncated space of excited states of the composite system.  The energies, spins, and parities of the target or of the composite nucleus, as well as the static (energy-independent) couplings  (\ref{eq:coupl_pot}), are  provided by a structure model of choice. However, we still need to compute the Green's function defined in Eq. (\ref{eq:Gmatrix}), i.e., the propagator in the space of excited states of the target. 

A possible way to obtain this propagator is based on its spectral decomposition (Lehmann representation), which involves a sum over the discrete spectrum of the restricted $Q$-space Hamiltonian and an integral over the continuum \cite{Dickhoff:05}. This approach requires the diagonalization of the propagator $G^{Q}$. In addition, one has to deal with the branch cut present in the continuum, which introduces a divergence in the integral for positive energies. This problem has been addressed in different ways, such as   by introducing a small but finite value in the ${\alpha}$ regularization parameter in (\ref{eq:Gmatrix}) \cite{BlanchonDAV2015}, or by making the spectral {decomposition} in  the Berggren basis \cite{RotureauDHNP2017}. 

Other approaches are based on providing an approximation for the potential defining the propagator. The simplest option is to assume that the nucleon is propagating freely in $Q$-space, in which case $G^{Q}$ is diagonal and equal to the free Green's function \cite{Slanina1969}. The authors of \cite{RaoRS1973} use instead the zero-order static potential $U_{00}$ to model the propagation of the nucleon in all the excited states, also assuming in this case that  the Green's function is diagonal. One of the advantages of this calculation with respect to the free propagation is to account for the effect of single-particle resonances (``shape''  resonances)  characteristic of the potential $U_{00}$ (see \cite{RaoRS1973}). In this study, we will take an additional step by  assuming that the intermediate system propagates according to the ``full'' optical potential $\mathcal V$ defined in (\ref{eq:OP}). 
{In other words,  the propagator in $Q$ space entering the
derivation of $\mathcal V$ is calculated by inverting the Hamiltonian derived from $\mathcal V$ itself.}

We will thus assume {a weak coupling approximation (see, \eg, Ref. \cite{Feshbach1958}),} that {is,} the $Q$-space propagator is diagonal and independent of the excited states, 
\begin{align}
G^{Q}_{ij}(E)=\widetilde G(E-\epsilon_{i})\delta_{ij}, \;(\text{for all }i,j),
\end{align}
 {which means} that the Green's function and the optical potential are connected by the two following equations,
 \begin{align}\label{eq102}
\nonumber \mathcal V(\mathbf{r,r'},E)&= U_{00}(\mathbf r)+\sum_{i \neq 0} U_{0i}(\mathbf{r})\widetilde G(\mathbf{r,r'},E-\epsilon_{i})U_{i0}
(\mathbf{r'})\\
 \widetilde G(\mathbf{r,r'},E)&=\lim_{{\alpha}\to0}\left(E-T- \mathcal V(\mathbf{r,r'},E)+i{\alpha}\right)^{-1},
\end{align}
 to be solved self-consistently. Since the optical potential is complex, the present approach accounts for absorption effects during the propagation in the virtually populated intermediate states. 
 
 The self-consistency of Eqs (\ref{eq102}) is enforced by the implementation of an iterative scheme, in which an iteration $\mathcal V^{(n+1)}$ of the optical potential is obtained from the previous $n$th iteration of the Green's function. Schematically, 
 \begin{align}
    &\mathcal V^{(0)}(E)=U_{00}, \nonumber \\
 \nonumber &\mathcal  V^{(1)}(E)=U_{00}+ \sum_i U_{0i}\,\widetilde G^{(0)}(E-\epsilon_{i})\, U_{i0}\\
    \vdots \nonumber \\
      &\mathcal  V^{(n+1)}(E)=U_{00}+ \sum_i U_{0i}\,\widetilde G^{(n)}(E-\epsilon_{i})\, U_{i0},
            \label{eq:iter}
\end{align}
where the Green's function is updated at every stage using 
 \begin{align}
 \widetilde G^{(n)}&(E-\epsilon_{i})=\lim_{{\alpha}\to0}\left(E-\epsilon_{i}-T- \mathcal V^{(n)}(E)+i{\alpha}\right)^{-1}.
\end{align}

This approach requires the development of a numerical procedure able to implement the above operator inversion for non-local potentials, which will be introduced in Sect. \ref{SIIB}.
It is to note that, although each intermediate state is propagated at the correct energy $E-\epsilon_{i}$, we make the approximation of always evaluating the optical potential at the same bombarding energy $E$. Although we have checked that the impact of this approximation is negligible after the first 2 iterations for some specific cases, a full understanding and validation of this procedure is still pending. The convergence of this iterative procedure is monitored by computing the volume integral of the optical potential
\begin{eqnarray}\label{volintegral}
    J^{(n)} = \int \mathcal V^{(n)}(\mathbf{r, r'})d\mathbf{r}d\mathbf{r'}
\end{eqnarray}
at each iteration for both the real and the imaginary part (see Fig. \ref{fig:24Mg_volint}).
\subsection{Calculation of the Green's function}\label{SIIB}
 The  partial wave decomposition of the Green's function can be calculated in a standard way as the product of the regular ($f_{l}$) and irregular ($g_{l}$) solutions of the radial wave equation  constructed with the optical potential (see, e.g., \cite{RaoRS1973,ArellanoB2021}),
\begin{align}
 \label{eq:Lehman_Wronskian}
 &\widetilde G_l(\mathbf r,\mathbf r',E)=
 \frac{\mu}{\hbar^2 r r'} \Big( \frac{f_l(\eta,k r_<)h_l^{(+)}(\eta,k r_>)}{\mathcal W(r_{<})} \nonumber \\
  &+\frac{f_l(\eta,k r_<)h_l^{(+)}(\eta,k r_>)}{ \mathcal W(r_{>})} \Big)\sum_m Y_l^m(\hat{r})Y_l^{m*}(\hat{r}')
\end{align} 
where $\hbar^{2} k^2/{2\mu}=E$ and $h^{(+)}_{l}=g_{l}+if_{l}$ has the asymptotic behavior of an outgoing Coulomb spherical wave. The radial coordinate $r_{<}$ ($r_{>}$) is the smallest (largest) among $r$ and $r'$.  The Sommerfeld parameter $\eta=\mu Z_t Z_p/\hbar^2k$ accounts for the effect of the Coulomb potential associated with the charges $Z_{t}$ and $Z_{p}$ of the target and projectile, and $\mu$ is the reduced mass of the system. The index $l$ labels the \emph{orbital} angular momentum, and the omission of the dependence of the Green's function on the total angular momentum $j$ essentially amounts to neglecting the effect of the spin-orbit (SO) term in the context of the present work. This term, included in most phenomenological OPs, primarily impacts spin observables; however, including SO could improve the description of backward‑angle cross sections. A microscopic SO implementation is a natural avenue for future work.
The above expression is a generalization with respect to the local potential case. More specifically, the Wronskian
\begin{align}
  \label{eq:Wronskian}
  \mathcal W=\frac{\partial f_l(\eta,kr)}{\partial r}g_l(\eta,kr)-\frac{\partial g_l(\eta,kr)}{\partial r}f_l(\eta,kr)
\end{align} 
corresponding to a  non local potential is not constant, in contrast with the Wronskian for a local potential, which has the constant value $\mathcal W=k$. 

We thus need to obtain the numerical solution to the radial integro-differential wave equation
\begin{align}\label{eq105}
\nonumber &\left(-\frac{\hbar^{2}}{2\mu}\frac{d^{2}}{dr^{2}}+V_{C}(\eta,r)+\frac{\hbar^{2} l(l+1)}{2\mu r^{2}}-E\right) c_{l}(kr)\\
&+\int \mathcal V_{l}(r,r',E)c_{l}(kr')\,d r'=0,
\end{align}
where $c_l\equiv f_l$ for regular and $c_l\equiv g_l$ for irregular solutions, $V_{C}$ is the Coulomb potential of a uniformly charged sphere (throughout this work, the Coulomb radius is the same as $U_{00}$ radius), and $\mathcal V_{l}$ are the terms of the multipolar expansion of the optical potential. For positive energies, the two linearly independent solutions satisfy the usual Coulomb asymptotic conditions in terms of the nuclear ($\delta_{l}$) and Coulomb ($\sigma_{l}$) phase shifts,
 \begin{align}\label{eq111}
    f_l(\eta,kr) \xrightarrow[r \to \infty]{} \sin(kr - l\pi/2-\eta \ln{(2kr)} + \sigma_l + \delta_l), \nonumber \\
    g_l(\eta,kr) \xrightarrow[r \to \infty]{} \cos(kr - l\pi/2-\eta \ln{(2kr)} + \sigma_l + \delta_l).
 \end{align}  
For negative energies, the asymptotic behavior correspond to Whittaker functions,
\begin{align}\label{eq110}
    \nonumber &F_l(\eta,\kappa r) \xrightarrow[r \to \infty]{} (2\kappa r)^{\eta}e^{\kappa r},\\
    &G_l(\eta,\kappa r)\xrightarrow[r \to \infty]{} (2\kappa r)^{-\eta}e^{-\kappa r},
\end{align}
with $\kappa=\sqrt{2\mu|E|}/\hbar$ and Sommerfeld parameter now defined as $\eta=\mu Z_t Z_p/\hbar^2\kappa$.

 To solve Eq. (\ref{eq105}) for the regular solution, we follow the procedure described in \cite{DescouvemontB2010}. The method is based on the inversion of the Hamiltonian matrix in the basis defined by the Lagrange functions
    \begin{align}\label{eq:reg_Lagrange}
      \varphi_i(r)=(-1)^{N+i}\left(\frac{r}{a x_i}\right)\sqrt{ax_i(1-x_i)}\frac{P_N(2r/a-1)}{r-ax_i}
    \end{align}
    defined in the interval $(0,a)$, where the $P_N(r)$ are Legendre polynomials, and $N$ is the number of elements in the basis, which are also the number of Lagrange points that constitute the radial grid. These points $x_{i}$ are the $N$ roots of the equation
\begin{align}
P_{N}(2x_{i}-1)=0.
\end{align}    
{Since these basis functions $\varphi_i$ are valid for $r \leq a$ only,
 the kinetic energy matrix is not Hermitian. To recover the Hermiticity of the kinetic energy operator in the finite interval $r \leq a$,} and to enforce {smooth} boundary conditions, the Hamiltonian is complemented with the Bloch operator \cite{blochFormulationUnifieeTheorie1957}, 
\begin{align}\label{eq:Bloch}
      \mathcal L_a=\frac{\hbar^2}{2\mu} \delta(r-a)\frac{d}{dr}.
\end{align}
This procedure for obtaining the regular solution for a non-local potential also provides the scattering phase shifts, obtained by matching with the appropriate asymptotic behavior. All the elastic scattering cross sections presented in this work have been obtained in this way.  
For a more detailed description of the numerical procedure, we refer to Ref. \cite{DescouvemontB2010}.

Since the irregular solution does not vanish at the origin, Eq. (\ref{eq105}) cannot be solved within the basis (\ref{eq:reg_Lagrange}), as all the basis functions satisfy $\varphi_i(0)=0$. We use instead the alternative basis
\begin{align}\label{eq:8}
      \tilde \varphi_i(r)=(-1)^{N+i}\sqrt{ax_i(1-x_i)}\frac{P_N(2r/a-1)}{r-ax_i}
\end{align}
defined in  an interval $(\epsilon,a)$, where $\epsilon>0$ is a small value of the radius, set to avoid the divergence of the irregular solution at $r=0$.

The numerical parameters used in this work are checked for the convergence of the results. In addition, the upper limit $a$ of the radial interval has to be larger than the range of the optical potential, so that the asymptotic expressions (\ref{eq111}) hold. In the context of the applications presented here,  we found $a=30$ fm as the size of the Lagrange radial box, and $N=100$ Lagrange points to be sufficient.

\section{Results and Discussions}

\subsection{{$p+^{40}$Ca and} $n+^{40}$Ca in a collective model}

\begin{figure} [t]
\centering
\includegraphics[width=0.49\textwidth]{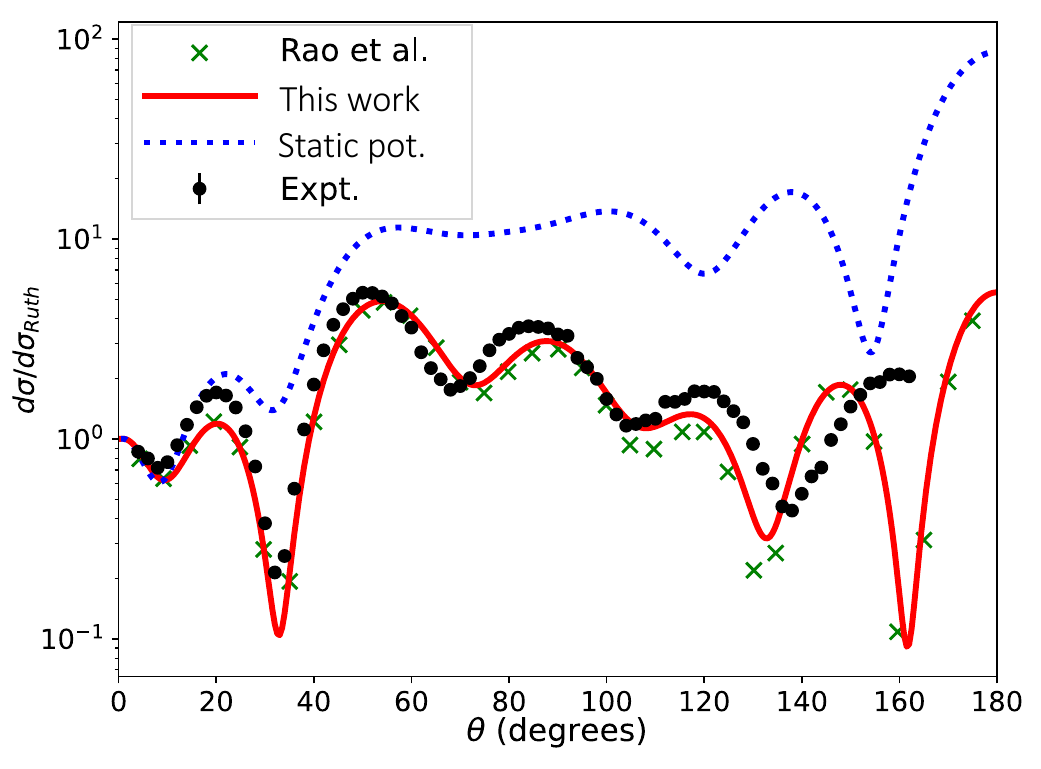}
\caption{ Angular differential cross sections {as
functions of the center-of-mass scattering angle for $^{40}$Ca$(p,p)$ at 30.3 MeV,} divided by Rutherford cross sections. The red curve is the calculation from our current work, the green crosses are the results of \cite{RaoRS1973}, and the black points correspond to experimental data from \cite{RidleyT1964}. The dotted blue line, which shows the calculation results employing only the static optical potential in Eq. (\ref{eq112}), demonstrates the impact of neglecting the coupling to the excited states. 
\label{fig:40Ca_p} 
} 
\end{figure}

{Let us momentarily depart in this section from the formalism described in Sect. \ref{SectIIA}, and benchmark the numerical methods described in Sect. \ref{SIIB}   against the results presented  in ref. \cite{RaoRS1973}. Instead of using a set of states of the composite (target plus nucleon) system, we follow this reference and} adopt the 10 vibrational states and couplings in  $^{40}$Ca presented in Table \ref{tab:40Ca_states} as the structure model for the  derivation of the $p+^{40}$Ca OP. {Here we also use the same states to derive} $n+^{40}$Ca OP {, not included in the original reference \cite{RaoRS1973}}. The coupling strengths are given in terms of $\beta_J$ deformation parameters that can be linked to the electromagnetic transition strengths from the excited states to the ground states, and  are extracted from $(\alpha,\alpha')$ measurements \cite{LippincottBA1967,Bernstein1969}.  The spin-orbit coupling effects are neglected. 

\begin{table*}[]
\caption{Excitation energies ($E_x$) and couplings for $^{40}$Ca vibrational states from Ref. \cite{RaoRS1973}. } \label{tab:40Ca_states}
\begin{tabular}{lllllllllll}
\hline
{J$^\pi$} & $1^-$ & $2^+$ & $2^+$ & $2^+$&  $3^-$ &  $3^-$ &$4^+$ & $4^+$ &  $5^-$ &  $5^-$ \\ \hline
$E_x$  (MeV)     & 18.0     & 3.9      & 8.0      & 16.0     & 3.73     & 15.73    & 8.0      & 20.0     & 4.48     & 16.48    \\
$\beta_J $     & 0.087    & 0.143    & 0.309    & 0.250    & 0.354    & 0.380    & 0.254    & 0.457    & 0.192    & 0.653    \\ \hline
\end{tabular}
\end{table*}

\begin{figure} [t]
\centering
\includegraphics[width=0.49\textwidth]{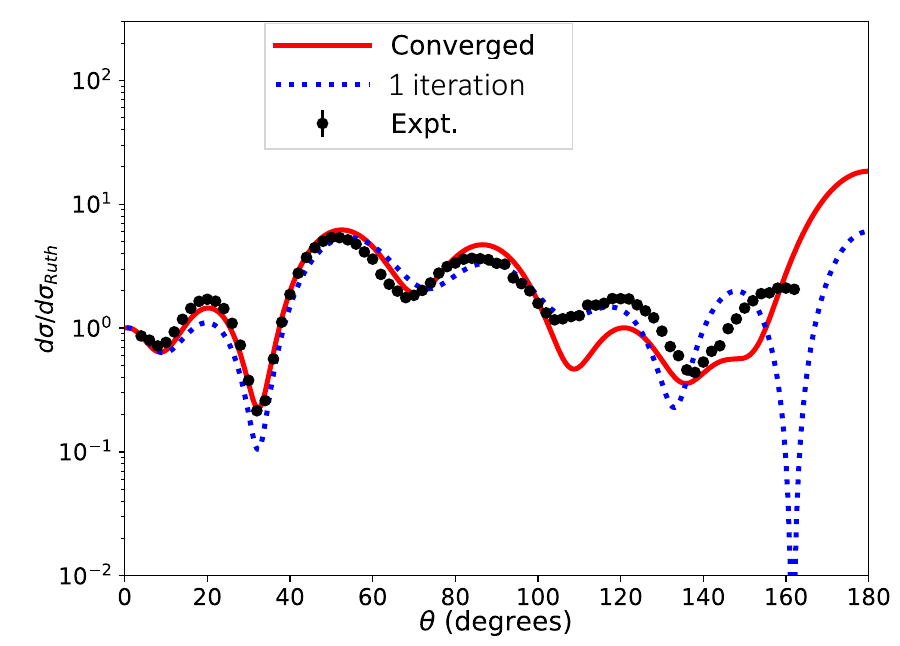}
\caption{ Angular differential cross sections {as functions of the center-of-mass scattering angle for $^{40}$Ca$(p,p)$ at 30.3 MeV,} divided by Rutherford using only the real part of the static potential (\ref{eq112}), for 1 (dotted blue line), and 30 (red line) iterations. 
\label{fig:40Ca_p_iter} 
} 
\end{figure}

More specifically, {in Ref. \cite{RaoRS1973}} the $U_{00}$ term of the optical potential is taken to be a local, Woods-Saxon potential with a  depth $U_0=50$ MeV, a diffuseness $a=0.75$ {fm}, and a radius $R_0=1.12 A^{1/3}$ fm. A small absorptive potential is added in terms of an imaginary potential with the same geometry and depth $W_0$=2 MeV.  The static, central potential can then be written as:
\begin{align}\label{eq112}
    U_{00}(r)=\frac{-(U_0+iW_0)}{1+\mathrm{exp}[(r-R_0)/a]}=U(r)+iW(r),
\end{align}
plus the Coulomb potential of a uniformly charged sphere of radius 1.12 $A^{1/3}$ fm. The couplings of the ground state with each of the 10 states (labeled by $i$) are modeled in terms of surface-peaked transition potentials,
\begin{align}
    U_{0i}(r)=(2\lambda+1)^{-1/2}\,r\,\frac{dU_{00}(r)}{dr}Y^{\mu*}_{\lambda}(\hat r),
\end{align}
where $\lambda,\mu$ are the angular momentum and its projection of the state $i$. Following Ref. \cite{RaoRS1973}, we use the weak coupling approximation, where
 the Green's function in the definition of the dynamic polarization potential (\ref{eq:100}) is diagonal.

Our benchmark is in good agreement with Ref. \cite{RaoRS1973} (Fig. \ref{fig:40Ca_p}), small deviations being likely due to numerical differences in the implementation that cannot be tracked down with the available information in the original reference {(also, cf. to Fig. 3 in Ref. \cite{RaoRS1973})}. The importance of the dynamic polarization potential is illustrated by the comparison with the calculation made with  $U_{00}$ alone (dotted blue line in Fig. \ref{fig:40Ca_p}), which yields significant deviations from the experimental data. In particular, the lack of absorption results in a large overestimation of the elastic cross section. The inclusion of all 10 collective states of $^{40}$Ca brings the calculation much closer to the measured points.  

The inclusion of a small imaginary part in the static potential (\ref{eq112}) {by the authors of Ref. \cite{RaoRS1973}}, introduces a non-dispersive element to the OP. 
We show in Fig. \ref{fig:40Ca_p_iter} a calculation without {this} imaginary {term}, $W(r)$, of the static potential {and find a reasonable agreement with the experimental data (blue dotted line)}. {In addition, we apply our iterative scheme (\ref{eq:iter}) onto this potential without $W(r)$. After iterating the potential and achieving convergence of the volume integral for each of the partial waves, we compare the cross sections obtained by computing the Green's function with the real part of the static potential (1 iteration), to the ones obtained after potential convergence has been reached (Converged).}
The cross sections at forward angles see a small change, getting closer to the experimental values. The effect of the iterations is more significant at backward angles, where the converged potential smooths out the last dip at $160^\circ$ (Fig. \ref{fig:40Ca_p_iter}). 

\begin{figure} [t]
\centering
\includegraphics[width=0.49\textwidth]{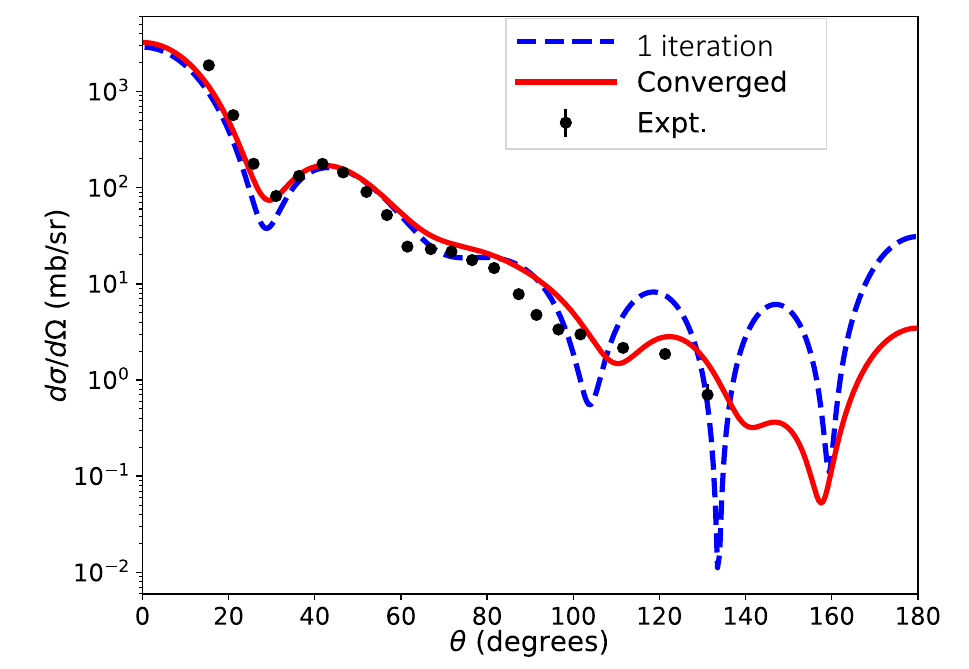}
\caption{ Angular differential cross sections as
functions of the center-of-mass scattering angle for $^{40}$Ca$(n,n)$ at 30.3 MeV. The blue dashed line corresponds to one iteration of the optical potential, and includes the phenomenological imaginary term $W(r)$ from \cite{RaoRS1973}, while the red curve shows results with converged OP (30 iterations) and without $W(r)$. 
The black points are experimental data from \cite{DeVitoASB1981}. 
\label{fig:40Ca_n} 
} 
\end{figure}

In addition to reproducing the proton elastic scattering presented in Ref. \cite{RaoRS1973}, we calculate the neutron elastic scattering cross sections for 30.3 MeV projectile energy using the same 10 vibrational states in $^{40}$Ca, by simply removing the Coulomb term from the static potential (Fig. \ref{fig:40Ca_n}). Similar to the proton scattering, the values at forward angles do not change drastically when the potential is iterated. However, there is a larger difference at $\theta > 90^\circ$ and the dips are smoother with the converged potential, which agrees better with the experimental data. These results highlight the impact of the iterative scheme (\ref{eq:iter}), showing the largest effect on scattering at backward angles. 

\subsection{$n+^{24}$Mg in the valence shell model}
{The inclusion of a phenomenological imaginary term in the static potential, as proposed by the authors of Ref. \cite{RaoRS1973}, serves to account for absorption processes that are not encompassed by the dynamic polarization potential}. They were thus addressing the fact that the reaction cross section for 30.3 MeV protons on $^{40}$Ca might not have been completely exhausted by the inelastic excitation of the 10 collective states present in their model. In principle, the implementation of the framework with an exhaustive description of the spectrum should mitigate the need for any added absorption, and should provide an overall better description of elastic scattering.  To explore this possibility,  we derive in this section an  $n+^{24}$Mg OP making use of the input provided by a $^{25}$Mg valence shell model calculations. 

The  spectrum of $^{25}$Mg  (around 600 states up to about 15 MeV of excitation energy) has been calculated using the PSDPF interaction \cite{BouhelalHCN2011}, built upon an inert $^4$He core. The valence nucleons occupy the \textit{p}, \textit{sd} and \textit{pf} shells, therefore allowing for the description of both positive and negative parity states. This choice of an extended model space allows for excitations from the $p$ to $sd$ shell, as well as the $sd$ to $pf$ shell, ensuring an exact separation of the center of mass coordinate.  In order to model the static potential $U_{00}(r)$, we use a simple real Woods-Saxon shape (\ref{eq112})  with parameters $U_0=48.5$ MeV, $R_0=3.69$ fm,  $a=0.65$ and $W_0=0.$ These values reproduce the experimental neutron separation energy of $^{25}$Mg by binding a $1d_{5/2}$ single particle state by 7.33 MeV. We note that changing these parameters within ~10\% does not significantly affect our final cross sections. 

We  now define the couplings of the incoming neutron with each one of the excited states of $^{25}$Mg. We first assume that the coupling matrix (\ref{eq:coupl_pot}) is diagonal.   The coupling potentials $U_{0i}(r)$ connecting the ground state to each excited state $i$ are taken to be of central real volume Woods-Saxon shape with the same radius and diffuseness as $U_{00}(r)$, but with a depth adjusted to reproduce the binding energy of each state. This choice of the coupling potential practically models each level $i$ as a single-particle state created by a mean field. These coupling potentials are then multiplied by a coupling strength taken to be the spectroscopic amplitude (overlap) $S_i$ of the corresponding state with the ground state of $^{24}$Mg. We then obtain an $l$-dependent OP, with the following radial part
\begin{align}\label{eq200}
 \nonumber \mathcal V_l({r,r'}&,E)= U_{00}(r)\\
 &+\sum_{\substack{i \neq 0\\\ell_i=l}}|S_i|^2 U_{0i}(r) \mathcal{G}_{\ell_i}({r,r'},E-\epsilon_{i})U_{i0}
(r'),
\end{align}
where $\ell_i$ is the orbital angular momentum of the state $i$, and
\begin{align}
  &\mathcal{G}_{\ell_i}(r,r',E)=
 \frac{\mu}{\hbar^2} \Big( \frac{f_{\ell_i}(\eta,k r_<)h_{\ell_i}^{(+)}(\eta,k r_>)}{\mathcal W(r_{<})} \nonumber \\
  &+\frac{f_{\ell_i}(\eta,k r_<)h_{\ell_i}^{(+)}(\eta,k r_>)}{ \mathcal W(r_{>})} \Big)   
\end{align}
is the radial part of the Green's function (\ref{eq:Lehman_Wronskian}).
The couplings being central, scattering partial waves only couple to $^{25}$Mg states with the same angular momentum. Since the shell model calculation used here  includes the \textit{p}, \textit{sd} and \textit{pf} shells, only the $\ell=0,1,2,$ and $3$  partial waves are involved in the calculation of the dynamic polarization potential.

With  this prescription, the contribution $\sigma_i$  of each state to the reaction cross section can be associated with the one-particle transfer cross section connecting the incoming scattering state with the corresponding final $^{25}$Mg state, expressed in the $prior$  representation (see, e.g., \cite{thompsonNuclearReactionsAstrophysics2009b}). 
This choice of the couplings might seem somehow heuristic, but, in this work, we explore the results obtained with this simple prescription, before taking a more microscopic approach  for the couplings in future works. In particular, a more microscopic systematic prescription would be to use the calculated transition densities. This is what has been done, for example, in \cite{BlanchonDAV2015}, albeit in a weak coupling approximation for a spectrum limited to the harmonic nuclear response (RPA).

\begin{figure} [ht]
\centering
\includegraphics[width=0.49\textwidth]{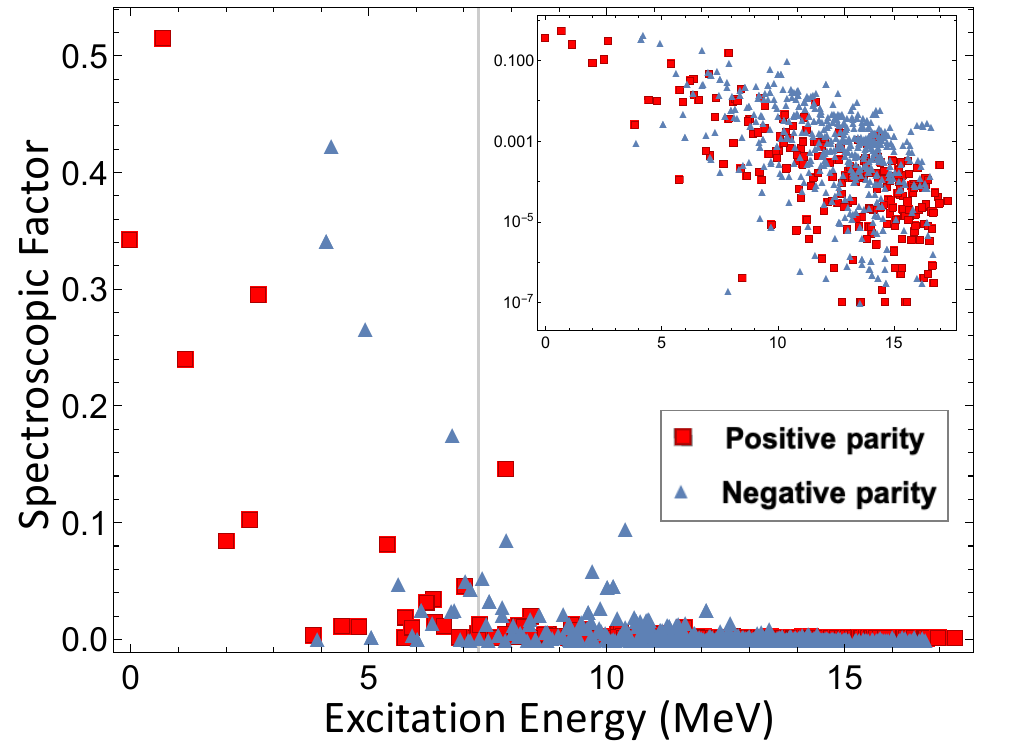}
\caption{\label{fig:24Mg_SF} Single-neutron spectroscopic factors of the $^{25}$Mg states  versus $^{25}$Mg excitation energy, calculated using the PSDPF interaction \cite{BouhelalHCN2011}. The gray vertical line shows the experimental neutron separation threshold in $^{25}$Mg. {Inset: same but in logarithmic scale}. 
} 
\end{figure}

We observe larger spectroscopic factors at the lower-lying positive and negative parity states (Fig. \ref{fig:24Mg_SF}), making their contribution the strongest in the construction of the optical potential. We note that, even though at higher energies the spectroscopic factors become orders of magnitude smaller (Fig. \ref{fig:24Mg_SF}, inset)  their contribution is not necessarily negligible as the density of states increases exponentially.

\begin{figure} [h]
\centering
\includegraphics[width=0.4\textwidth]{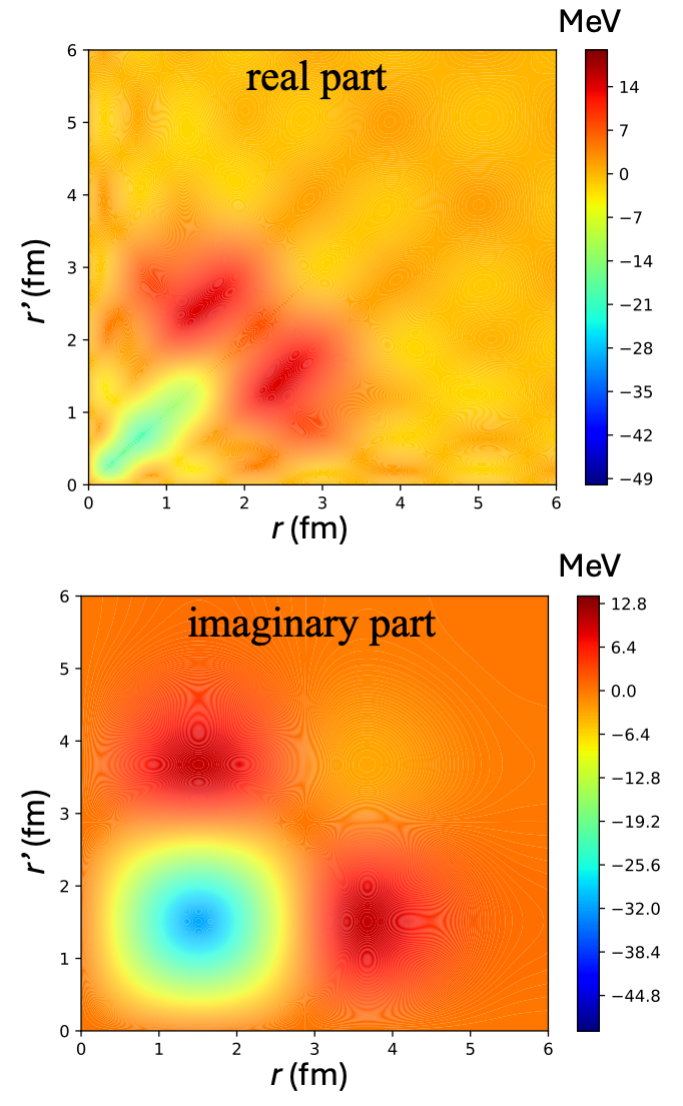}
\caption{\label{fig:24_pot} Real and imaginary parts of the $n+^{24}$Mg dynamic polarization potential (see Eq. \ref{eq:100}) for  $E=$3.4 MeV, calculated  for the  $\ell=1$ partial wave. 
} 
\end{figure}

The resulting energy-dependent dynamic polarization potential, $V_\mathrm{dpp}$, is non-local in nature and has both real and imaginary terms (Fig. \ref{fig:24_pot}). We calculate it at each neutron bombarding energy and add the static local $U_{00}$ term to obtain the full optical potential. By examining the values of $V_\mathrm{dpp}$ one can see its significant contribution in the calculations of scattering observables as the lowest values of $V_\mathrm{dpp}$ are comparable to the depth of $U_{00}$ Woods-Saxon potential. The periodicity of $V_\mathrm{dpp}$ with respect to $r$ and $r'$ is related to the periodicity of the regular and irregular solutions that enter in the expression of the Green's function in Eq. (\ref{eq:Lehman_Wronskian}). We note that the imaginary part of the potential has both positive and negative values (Fig. \ref{fig:24_pot}, bottom panel), however its volume integral always remains negative as it cannot create any flux.  


\begin{figure} [h]
\centering
\includegraphics[width=0.49\textwidth]{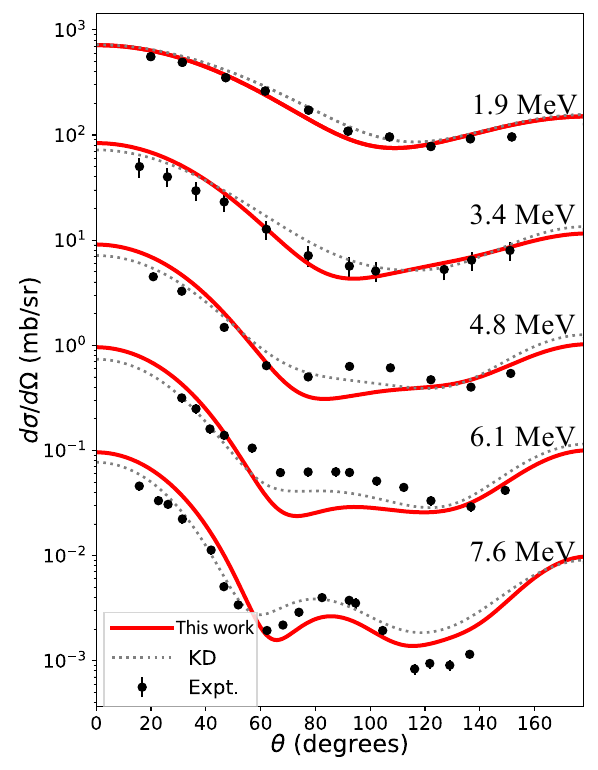}
\caption{\label{fig:24Mg_all} Angular differential cross sections as
functions of the center-of-mass scattering angle for $^{24}$Mg$(n,n)$ at 5 different incident energies labeled next to each curve. The red curves correspond to the microscopic OP in Eq. (\ref{eq200}), while the dotted lines have been obtained with the Koning-Delaroche (KD) global optical potential \cite{KoningD2003}. The curves and data points at the top represent true values, while the others are offset by factors of 10, 100, etc. 
The experimental data (black dots) are part of the body of data used to fit the KD phenomenological potential.
} 
\end{figure}

Our main results are presented in Fig. \ref{fig:24Mg_all}. We show angular differential elastic scattering $n+ ^{24}$Mg cross sections for 5 different incident energies. In our calculations, all processes contributing to the formation of an excited $^{25}$Mg compound nucleus state contribute to the imaginary part of the optical potential, and thus to the reaction cross section and the depletion of the elastic channel. A fraction of these compound nucleus states can decay emitting a neutron with an energy equal to the incident one, contributing to the compound elastic cross section. Since these events cannot be experimentally distinguished from direct elastic scattering,  we have added incoherently to our calculations the compound elastic contribution calculated with the Hauser-Feshbach code  \texttt{YAHFC} \cite{Ormand2021YAHFC}. The resulting comparison with experimental data, as well as with the curves obtained with the Koning-Delaroche global optical potential \cite{KoningD2003} are remarkable given that our calculations contain no phenomenological imaginary terms. 
In order to be completely consistent, the Hauser-Feshbach calculation should have been performed using our optical potential  instead of the default Koning-Delaroche used in \texttt{YAHFC}. We leave for future works a more consistent treatment of the neutron decay following the compound nucleus formation.

\begin{figure} [t]
\centering
\includegraphics[width=0.48\textwidth]{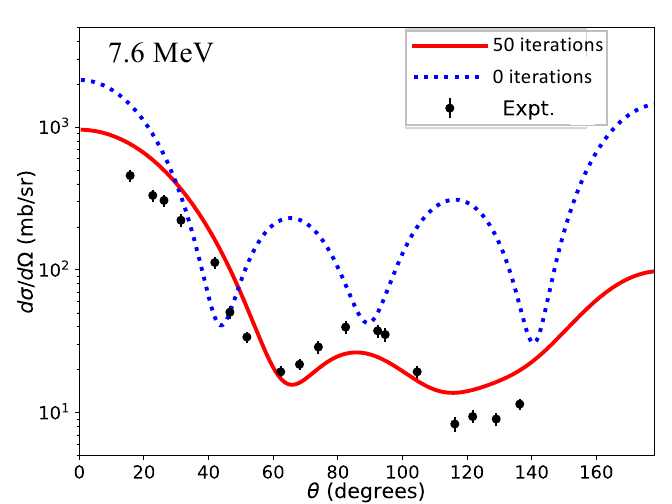}
\caption{\label{fig:24Mg_7_6} Angular differential cross sections as
functions of the center-of-mass scattering angle for $^{24}$Mg$(n,n)$ at 7.6 MeV. The dotted blue line corresponds to the calculation with only the $U_{00}$ term of the optical potential in Eq. (\ref{eq200}). 
} 
\end{figure}

We illustrate the role of the dynamic polarization potential by comparing the cross section obtained with the full optical potential (\ref{eq200}) to the one making use of the static potential $U_{00}(r)$ only (Fig. \ref{fig:24Mg_7_6}). The effects of the absorption introduced by the coupling to the many-body states of $^{25}$Mg, as well as the renormalization of the real part of the optical potential associated with the virtual population of these states, is very apparent, and brings the calculated elastic scattering cross section much closer to the experimental values.

An important aspect of a unified description of structure and reactions is the treatment of the continuum. In this  work, the structure of $^{25}$Mg has been calculated in a harmonic oscillator basis, so the computed spectrum is fully discrete and bound, even above the experimental neutron separation energy of 7.33 MeV. However, in our approach we interpret the shell model states above the neutron separation energy as resonances, and the Green's function $\widetilde G(E)$ exhibits the correct asymptotic behavior as a function of the energy, determined by the corresponding irregular solution $g_l$ in (\ref{eq111}). Therefore, an arbitrary scattering state obtained acting on a free wave $\phi_0$ with the M\"oller operator $\Omega(E)$
\begin{align}
\Psi(E)=\Omega(E)\phi_0(E)=\left(1+\widetilde G(E)\mathcal V(E)\right)\phi_0(E),
\end{align}
will exhibit the desired resonance behavior at the poles of the Green's function, with widths determined by the penetrability factors {and} the corresponding spectroscopic factors. This approach, in which the discrete shell model spectrum above the particle emission threshold is interpreted as many-body resonances that are coupled to the continuum, is reminiscent of the continuum shell model formalism \cite{rotterContinuumShellModel1991}. 

{Let us briefly comment on the important issue of the fragmentation of doorway states and giant resonances associated with their damping width \cite{bertschDampingNuclearExcitations1983,bortignonGiantResonances1998}, also in connection with the energy-averaging of the OP (see e.g. \cite{brownFoundationsOpticalModel1959,blockNeutronStrengthFunction1963,MahauxBBD1985}). Let us first stress that the potential (\ref{eq200}), when evaluated at a real value of the energy $E$, is not an energy-averaged quantity, thus exhibiting the rapid energy variation associated with the poles of the Green's function. On the other hand, our shell-model calculated number of levels (around 50 up to the neutron separation energy) matches the experimental one, suggesting that we capture reasonably well the fragmentation of the doorway states. Since the coupling potentials are real, the entire absorptive imaginary part of our OP is due to the coupling of the ground state of $^{24}$Mg to the shell model states of $^{25}$Mg. In this context, the energy-averaged OP usually associated with Optical Model could be obtained by calculating (\ref{eq200}) with a small imaginary component added to the level energies.}

\begin{figure} [t]
\centering
\includegraphics[width=0.5\textwidth]{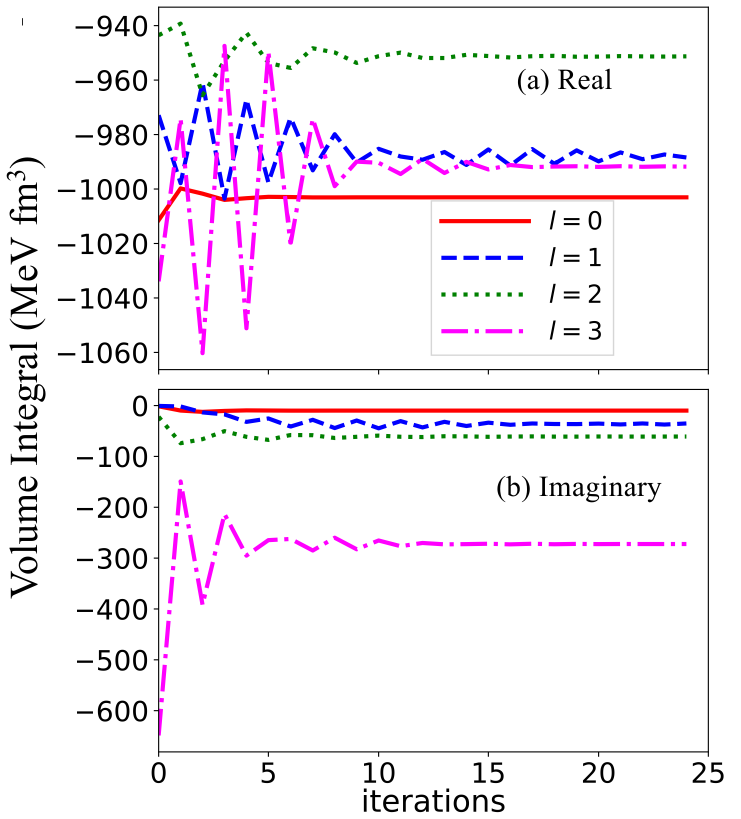}
\caption{\label{fig:24Mg_volint} Volume integral convergence of $^{24}$Mg OP for 7.6 MeV vs. the number of iterations for $l=0$ to 3 partial waves for the (a) real and (b) imaginary parts of the potential. 
}
\end{figure}

 We now turn our  attention to the convergence of the iterative scheme described in (\ref{eq:iter}) by computing the volume integral (\ref{volintegral}) for each relevant partial  wave as a function of the iteration number. For of 7.6 MeV incident neutrons,  we observe that the calculations are fully converged after 15-20 iterations for all the partial waves, and both for the real and the imaginary part of the optical potential (Fig. \ref{fig:24Mg_volint}).

\begin{figure} [h]
\centering
\includegraphics[width=0.48\textwidth]{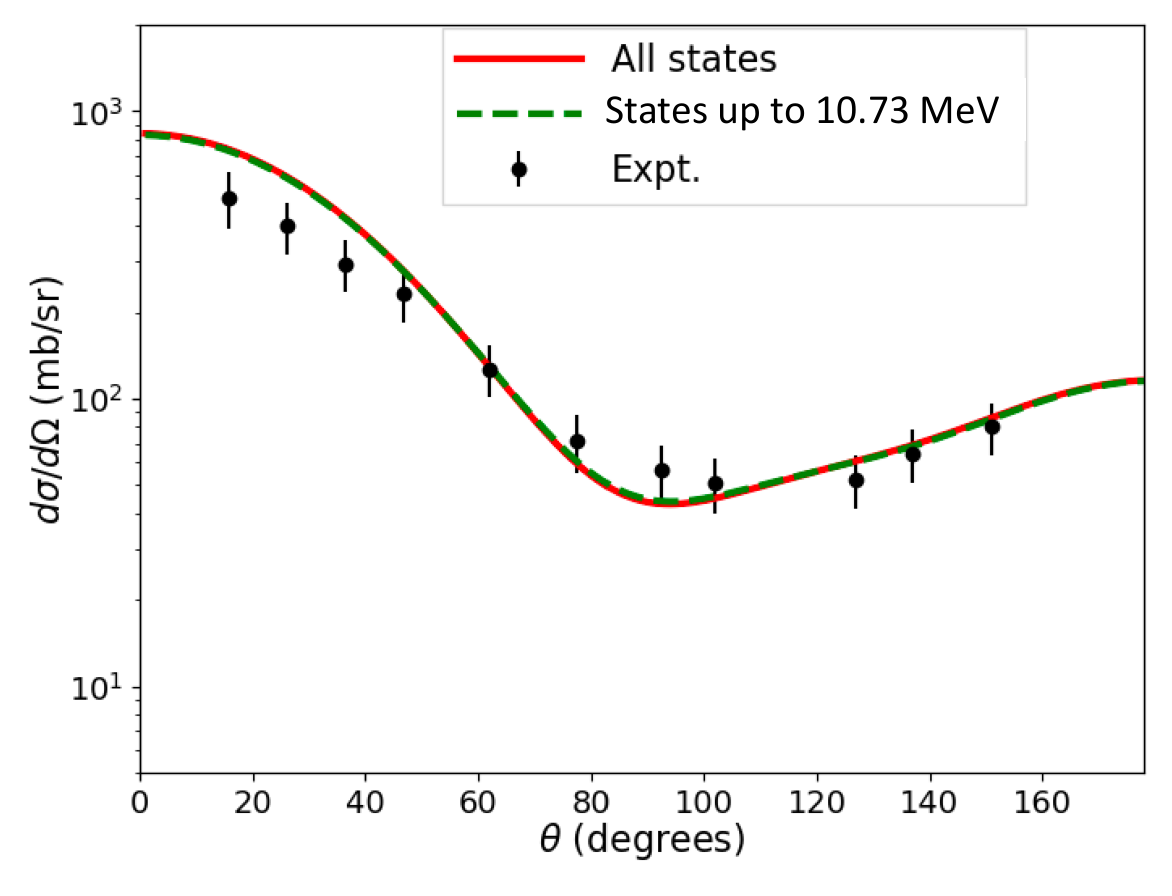}
\caption{\label{fig:24Mg_neg+e} Angular differential cross sections as functions of the center-of-mass scattering angle for $^{24}$Mg$(n,n)$ at 3.4 MeV. The dashed green line corresponds to the optical potential constructed using only states up to 10.73 MeV in the excitation energy of $^{25}$Mg (corresponding to the bombarding energy plus the 7.33 MeV neutron separation energy of $^{25}$Mg).
} 
\end{figure}

An important aspect of the numerical convergence of the calculation is associated with the truncation of the Hilbert space. Aside from the consideration of the total number of states provided by the shell model calculation, the problem at hand offers a natural distinction between states situated above and below the considered bombarding energy. Although most calculations to date neglect the contribution of states above the bombarding energy, their effect in the renormalization of the real part of the optical potential should be addressed. These states in the general expression (\ref{eq200}) correspond to negative values of the argument $E-\epsilon_i$ of the Green's function, which then acquires an exponentially decaying asymptotic behavior (see Eq. (\ref{eq110})). We show in Fig. \ref{fig:24Mg_neg+e} the elastic scattering cross section for 3.4 MeV incident neutrons, calculated with and without the inclusion of states above the bombarding energy. The results are almost identical, which supports the notion that not including them in the calculation is a reasonable approximation.

\section{Conclusions and outlook}
We have presented a new formalism inspired by the Feshbach theory for the derivation of optical potentials from microscopic structure input. The approach is based on the explicit calculation of the contribution of each excited state of the composite (nucleon+target) system, making use of the single-particle Green's function (propagator). It is then, in principle, able to accommodate for any structure input providing the spectrum of the compound nucleus and the coupling potentials, as well as the overlaps with the elastic channel. In particular, single-nucleon overlaps have been calculated in different  \emph{ab initio} models \cite{NavratilBC2006,BridaPW2011,McCrackenNMQH2021,SargsyanLSMD2023} and could be used within the present approach. More generally, the range of validity of the calculations in terms of masses and bombarding energies is  determined by the specific approach used to describe the structure of the composite system. The resulting optical potential is non-local, complex, dispersive, and depends on both the energy and the angular momentum. 

In order to model the propagator, we develop a numerical method to calculate the Green's function for non-local potentials, based on a modification of the $R$-matrix approach presented in \cite{DescouvemontB2010} which allows for the calculation of the irregular solutions of the corresponding wave equation. We are then able to impose self-consistency between the optical potential and the Green's function by implementing an iterative procedure.  

We apply our formalism to the construction of the n+$^{24}$Mg OP using the states of the composite $^{25}$Mg nucleus calculated by shell model. The neutron elastic scattering cross sections for $^{24}$Mg using this OP exhibit good agreement with the experimental data for several bombarding energies. The results obtained with the shell model description of $^{25}$Mg are encouraging enough to motivate the implementation of several improvements over the present work. More specifically, we plan in the near future to improve upon the choice of coupling potentials (\ref{eq:coupl_pot}).
In particular, it is possible to use a non-local static potential  obtained from a Hartree-Fock calculation, while the non-diagonal couplings associated with the connection of the ground state with the excited states can be obtained from a microscopic calculation of the corresponding transition densities. 

Once the structure calculation has been made available, the derivation of the optical potential is not computationally expensive, at least for a number of states not exceeding a few thousands. This opens the door for exploring the onset of the statistical regime as the density of states increases by calculating explicitly the energy-averaged optical potential, as well as the fluctuations with respect to the average value.  According to the standard interpretation of the Optical Model, the latter is associated with compound processes, while the former describes direct reactions \cite{feshbachModelNuclearReactions1954,brownFoundationsOpticalModel1959,blochFormulationUnifieeTheorie1957}. We hope that research along these lines could shed light on the interplay between direct and compound reactions in general. More specifically, this would likely allow for the exploration of the limits of validity of  Hauser-Feshbach theory as we address nuclei further from stability, for which the level density is expected to drop significantly. This region of the nuclear chart is of great importance for astrophysics and, more generally, for applications where unstable fission fragments are present.    

\section*{ACKNOWLEDGMENTS}
We thank G. Blanchon and C. Hebborn for useful discussions and C. Pruitt for providing compound elastic contributions. We are also grateful to the anonymous referee for the detailed review and invaluable comments, which helped improve the current manuscript. This work was performed under the auspices of the U.S. Department of Energy by Lawrence Livermore National Laboratory under Contract DE-AC52-07NA27344, with support from LDRD project 21-ERD-006. This material is based upon work supported by the U.S. Department of Energy, Office of Science, Office
of Nuclear Physics, under the FRIB Theory Alliance award DE-SC0013617.

\bibliographystyle{apsrev}
\bibliography{microop}

\end{document}